\documentclass[10pt,aps,prl,superscriptaddress,showpacs,preprint,floatfix]{revtex4-1}

\usepackage{amsmath}
\usepackage{amssymb}
\usepackage{graphicx}

\begin{document}

\title{Unoccupied Topological States on Bismuth Chalcogenides}

\author{D. Niesner}
\author{Th. Fauster}
 \affiliation{Lehrstuhl f\"ur Festk\"orperphysik, Universit\"at Erlangen-N\"urnberg, D-91058 Erlangen, Germany}

\author{S.~V. Eremeev}
 \affiliation{Institute of Strength Physics and Materials Science,
634021, Tomsk, Russia}
 \affiliation{Tomsk State University, 634050 Tomsk, Russia}

\author{T.~V. Menshchikova}
 \affiliation{Tomsk State University, 634050 Tomsk, Russia}

\author{Yu.~M. Koroteev}
 \affiliation{Institute of Strength Physics and Materials Science,
634021, Tomsk, Russia}
 \affiliation{Tomsk State University, 634050 Tomsk, Russia}

\author{A.~P. Protogenov}
 \affiliation{Institute of Applied Physics, Nizhny Novgorod, 603950,
Russia}
 \affiliation{ Donostia International Physics Center
(DIPC), 20018 San Sebasti\'an/Donostia, Basque Country, Spain}

\author{E.~V. Chulkov}
 \affiliation{ Donostia International Physics Center
(DIPC), 20018 San Sebasti\'an/Donostia, Basque Country, Spain}
 \affiliation{Departamento de F\'{\i}sica de Materiales UPV/EHU and
Centro de F\'{\i}sica de Materiales CFM and Centro Mixto CSIC-UPV/EHU,
20080 San Sebasti\'an/Donostia, Basque Country, Spain}

\author{O.~E. Tereshchenko}
 \affiliation{Institute of Semiconductor Physics, Novosibirsk, 630090 Russia}

\author{K.~A. Kokh}
 \affiliation{Institute of Geology and Mineralogy, Novosibirsk, 630090 Russia}

\author{O. Alekperov}
\author{A. Nadjafov}
\author{N. Mamedov}
 \affiliation{Institute of Physics, Azerbaijan National Academy of Sciences, AZ1143 Baku, Azerbaijan}

\begin{abstract}
The unoccupied part of the band structure of topological insulators
Bi$_2$Te$_{x}$Se$_{3-x}$ ($x=0,2,3$) is studied by angle-resolved
two-photon photoemission and density functional theory.  For all surfaces linearly-dispersing surface states are
found at the center of the surface Brillouin zone at energies around 1.3\,eV above the Fermi level.  Theoretical
analysis shows that this feature appears in a spin-orbit-interaction
induced and inverted local energy gap.
This inversion is insensitive to variation of electronic and structural parameters in Bi$_2$Se$_3$ and Bi$_2$Te$_2$Se. In Bi$_2$Te$_3$ small structural variations can change the character of the local energy gap depending on which an unoccupied Dirac state does or does not exist.
Circular dichroism measurements confirm the expected spin texture.  From these findings we
assign the observed state to an unoccupied topological surface state.
\end{abstract}

\pacs{73.20.-r, 79.60.Bm}

\maketitle

Three-dimensional topological insulators (TIs) are insulators
in bulk and metals at the surface. Metallic character of a TI
surface is determined by a massless Dirac state that crosses the
Fermi level $E_{\mathrm F}$ \cite{FKM_2007,Qi_2008,Zhang_NatPhys2009}.
This spin-polarized linearly-dispersing surface state arises from a
symmetry inversion of the bulk bands at band gap edges owing to
strong spin-orbit interaction (SOI). Various new phenomena in
TIs were predicted like dissipationless spin transport
\cite{Roushan_Nat2009}, formation of Majorana fermions in the presence
of superconductors \cite{FuKane2008} and magnetic monopoles
\cite{Qi27022009}.

Most of these phenomena deal with low-energy electronic excitations and
transport.  Little experimental work has been performed on excited
electronic states and their dynamics.  The best studied
three-dimensional topological insulators at present are bismuth
chalcogenides.  The equilibrium band structure of these TI's is well understood
\cite{Zhang_NatPhys2009,Chen10072009,Hsieh09,xia_2009,Yazyev2010,EremeevJETPL2010,Kuroda2010,bianchi2010}
and the spin texture of the topological surface state (TSS) was observed
both indirectly by use of circular dichroism (CD) \cite{Wang2011} and
directly by spin-resolved experiments \cite{Hsieh09,Eremeev_NatComm}.

The materials are usually intrinsically $n$-doped, which complicates
direct spectroscopic access to excited electrons in a TSS\@. Besides
an earlier inverse photoemission investigation \cite{ueda1999} a
recent study was performed by Sobota {\it et al.} demonstrating that
a persistent occupation of the Dirac cone close to the Fermi level
in $p$-doped Bi$_2$Se$_3$ can be supported even picoseconds after an
initial optical excitation \cite{PhysRevLett.108.117403}.

Here we present a different approach by investigating higher-excited
TSSs in Bi$_2$Te$_{x}$Se$_{3-x}$ ($x=0,2,3$).  In the calculated band
structures symmetry-inverted band gaps are identified which
support an empty Dirac cone.  Experimental proof is given by
monochromatic two-photon photoemission (2PPE).  Calculations together
with CD measurements reveal the characteristic spin structure of a
topological surface state.

Angle-resolved monochromatic 2PPE and one-photon photoemission
(ARPES) experiments were conducted using the third and fourth
harmonic (4.65\,eV and 6.2\,eV photon energy) of a titanium:sapphire
oscillator with a repetition rate of 80\,MHz and pulse lengths around 100\,fs
\cite{Boger05njp,suppmat}.  In both
cases the beam is initially $p$-polarized with an incidence angle of
45$^\circ$.  Circular polarization of the third harmonic necessary for
circular dichroism experiments was obtained using a $\lambda
/4$-waveplate.  Two-dimensional momentum distribution patterns (MDCs) at
constant kinetic energy were recorded using an ellipsoidal
``display-type" analyzer at an energy and angular resolution of 50\,meV and 3$^\circ$, respectively \cite{rieg83,suppmat}. Angle-resolved spectra for the intensity maps were acquired by a hemispherical analyzer
with resolution of 34\,meV and 1.6$^\circ$, respectively \cite{Boger05njp}.

Bi$_2$Te$_2$Se was prepared from presynthesized mixtures of
Bi$_2$Te$_3$ and Bi$_2$Se$_3$, which in turn were prepared from
elementary Bi, Te and Se of 99.999\% purity. Crystal growth for all
three compounds was done in sealed quartz ampoules coated with
a carbon layer.  For recrystallization we used a vertical variant of the
modified Bridgman method \cite{Kokh05}.  The resulting ingots consisted
of one or several large single-crystalline blocks.
{Bi$_2$Se$_3$, Bi$_2$Te$_3$ and Bi$_2$Te$_2$Se samples were naturally $n$-doped (through the formation of defects) with carrier concentration in the range of $(1-9)\times 10^{18}$ cm$^{-3}$.}
Samples were cleaved
in vacuum at a pressure $<5\times10^{-6}$~Pa and then transferred to
ultrahigh vacuum (pressure $<1\times10^{-8}$~Pa) where they were cooled
to 90\,K for measurements.  Sample quality and orientation was checked
by low-energy electron diffraction (LEED) showing a sharp threefold pattern.  According to LEED patterns the samples were oriented with
the laser beams incident parallel to the $\overline\Gamma\overline{\mathrm M}$ mirror plane or along the $\overline\Gamma\overline{\mathrm K}$ direction.

\begin{figure}
        \includegraphics[width=.88\columnwidth]{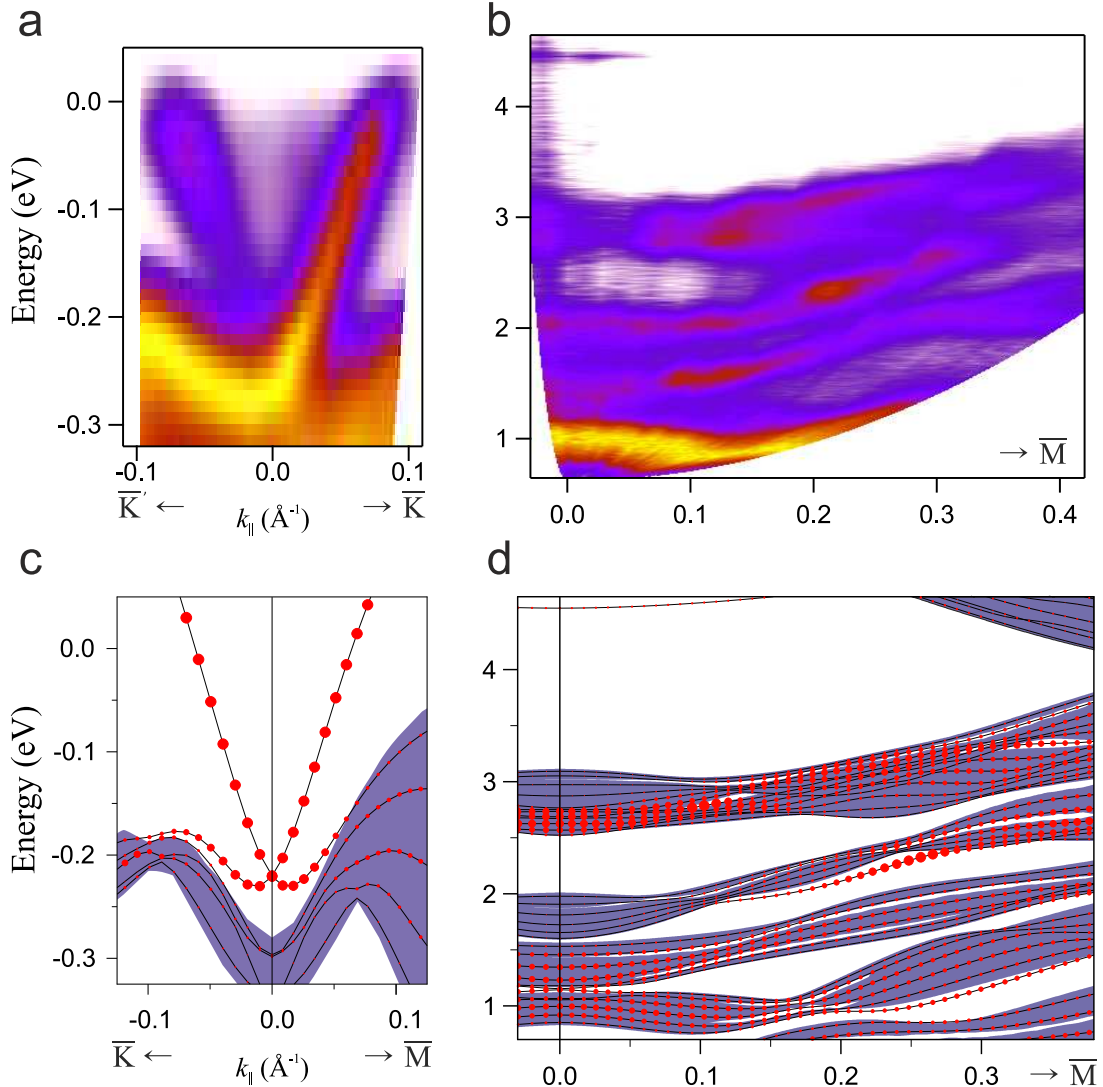}
    \caption{\label{fig1}(Color online) Measured (top row) and
calculated (bottom row) data on the band structure of
Bi$_2$Te$_2$Se. {Left (right) column shows occupied (unoccupied) states. Shaded regions in the calculations, (c) and (d), indicate
projected bulk bands, whereas red dots indicate an increased
localization within the topmost QL}.}
\end{figure}

Because the description of the unoccupied states is a delicate
issue of density functional theory (DFT) we employ two different
computer codes that are based on DFT\@. The main part of the
electronic structure calculations was performed within the density
functional formalism implemented in VASP \cite{VASP1,VASP2}. We used
the all-electron projector augmented wave (PAW) \cite{PAW1,PAW2}
basis sets with the generalized gradient approximation of Perdew,
Burke, and Ernzerhof (PBE) \cite{PBE} to the exchange correlation
(XC) potential. For bulk band structure calculations the local
density approximation (LDA) \cite{LDA} to XC potential was also tested.
The experimental lattice parameters were used for calculation
of all the considered compounds while atom positions within the unit cell were optimized.
The Hamiltonian contains scalar relativistic corrections, and the
SOI is taken into account by the second variation method \cite{KH}.
To simulate the semi-infinite Bi$_2$Te$_{x}$Se$_{3-x}$(111) we use a
slab composed of 9 quintuple layers (QLs).

The second approach used for electronic structure calculations is
the full-potential linearized augmented plane-wave (FLAPW) method as
implemented in the FLEUR code \cite{FLAPW} with PBE for the
exchange-correlation potential. Spin-orbit coupling was included in
the self-consistent calculations as described in Ref.~\cite{Li}. The
FLAPW basis has been extended by conventional local orbitals to
treat quite shallow semi-core $d$-states. Additionally, to describe
high-lying unoccupied states accurately, we have included for each
atom one local orbital per angular momentum up to $l$ = 3. The
fundamental quantities are consistently reproduced by both methods \cite{suppvasp}
except for the case Bi$_2$Te$_3$ which will be addressed below in
detail.

To study the effect of dispersion interactions we use the van-der-Waals 
non-local correlation functional (vdW-DF) as implemented
in the VASP code \cite{[][. The optB86b-vdW functional was used in the present work.]Klimes}.
Both lattice parameters and internal atomic positions were optimized in this approach.

\begin{figure}
        \includegraphics[width=.88\columnwidth]{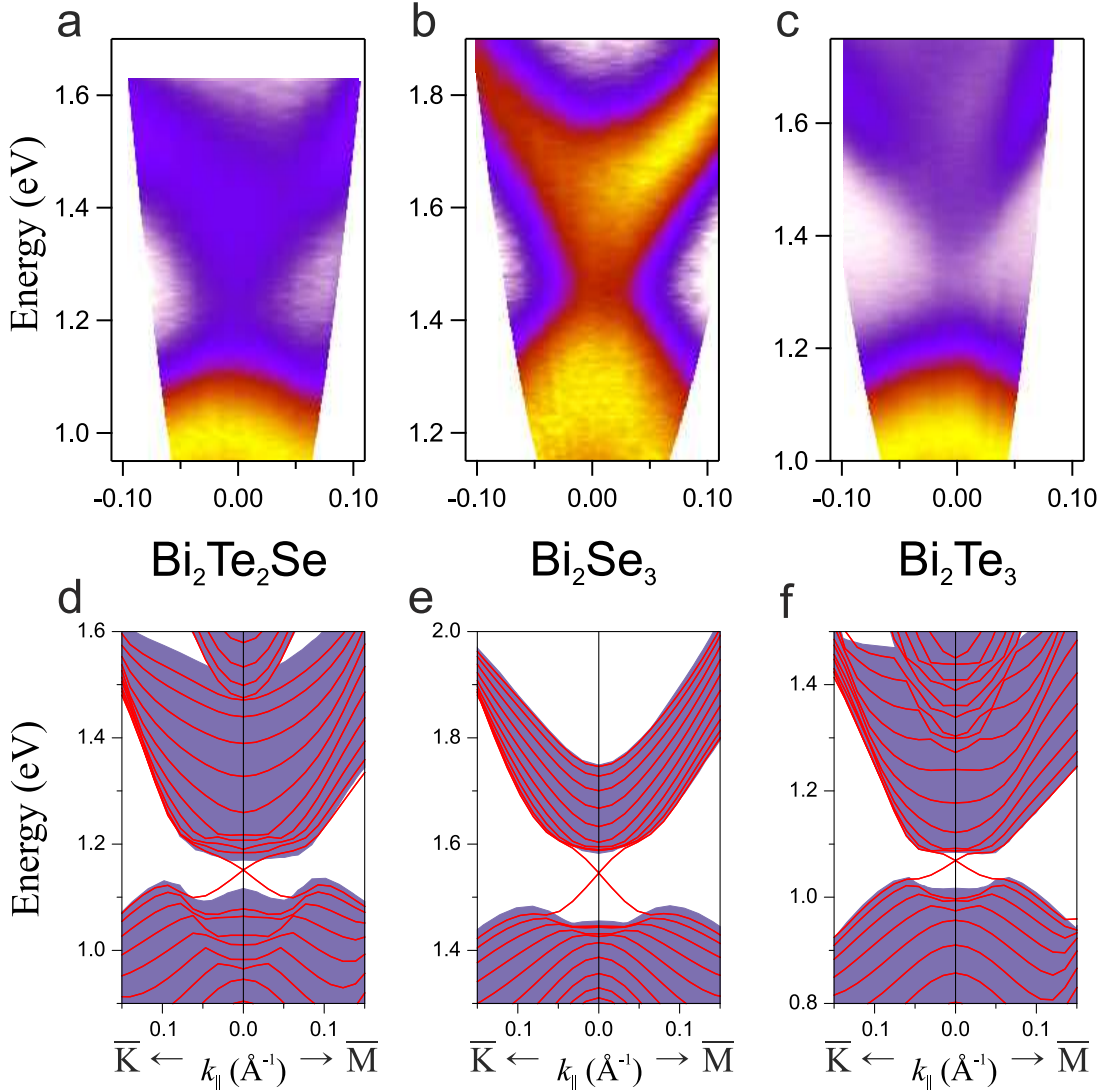}
\caption{\label{fig2}(Color online) Close-up of the band gap under
investigation for the three materials. No azimuthal dependence was observed close to 
$\overline{\Gamma}$ and the sample orientation was optimized to access the Dirac point in experiment.}
\end{figure}

We start with the ternary tetradymite compound Bi$_2$Te$_2$Se which
recently was shown to be a three-dimensional TI by ARPES measurements
\cite{Neupane2012,Miyamoto2012}.  The crystal structure of this compound is
obtained by replacing the central Te layer of each QL of Bi$_2$Te$_3$ by
a Se layer.  The results of photoemission experiments and DFT calculations on Bi$_2$Te$_2$Se are shown in
Fig.~\ref{fig1}.  Shaded regions in the calculated band structure depict
projected bulk bands, while red dots indicate an increased localization
in the first quintuple Bi$_2$Te$_2$Se layer.  From the ARPES data (Fig.~\ref{fig1}(a)) the
Dirac cone in the occupied part of the band structure is found
$0.25$\,eV below $E_{\mathrm F}$. Depending on crystal growth conditions values between 0.1 and 0.4~eV have been found \cite{Neupane2012,Miyamoto2012}.
In the 2PPE data (Fig.~\ref{fig1}(b)) all
prominent features can be attributed to unoccupied bands in the
calculations (Fig.~\ref{fig1}(d)) and we conclude that the 2PPE process
is dominated by intermediate states.  The measured bands of high
intensity coincide well with surface resonances in the calculation,
giving evidence of a high surface sensitivity.  A similar degree of
agreement between theory and experiment is also found for Bi$_2$Se$_3$
and Bi$_2$Te$_3$ surfaces. The present results are in qualitative agreement with
the inverse photoemission spectra which show three conduction band peaks
\cite{ueda1999}.

In the following we will focus on the projected bulk band gaps at
energies above $E_{\mathrm F}$.  Figures~\ref{fig2}(a) and (d) show
a close-up of such a gap for Bi$_2$Te$_2$Se.  In fact both 2PPE and
DFT data show that it is bridged by {two linearly dispersing
bands which cross} at the $\overline{\Gamma}$-point. At this energy it appears in MDCs as a single spot evolving
into a circle towards higher energies (see also Fig.~\ref{fig4}(a) and \cite{suppmdcs}).
At energies below the crossing point it turns into a surface resonance degenerate with
bulk bands.

The appearance of the Dirac state in the conduction-band local gap
raises the question:  Is this surface state topologically protected?  In
contrast to the Dirac state located in the principal energy gap the
new massless Dirac state is located in an local energy gap and upon going
along a path between time-reversal invariant momenta (TRIMs) in the
Brillouin zone the gap in the spectrum of bulk states closes and opens.

The origin of the reduction of the contribution made by extended
states can be found by comparing the continuous and lattice versions
of the $\mathbb{Z}_2$ invariant. Analysis of the relation
$(-1)^{\nu_{0}}=(-1)^{2P_{3}}$ \cite{WQZ} of the index $\nu_{0}$
\cite{FKM_2007} in the theory of topological insulators with the
winding number $2P_{3}$ does not solve the problem. It is shown in
Ref.~\cite{FK} that the $\mathbb{Z}_2$ invariant in the continuous
case is alternatively expressed as
\begin{alignat}{1}
D&=\frac{1}{2\pi i}\left[ \oint_{\partial{\cal B}^-}A-\int_{{\cal
B}^-} F \right]\,\,\, {\text{\rm mod 2} }, \label{FuKane}
\end{alignat}
where ${\cal B}^-=[-\pi,\pi]\otimes[-\pi,0]$ is half of the
Brillouin zone, $A=\mathrm{Tr} \, \psi^{\dagger}d\psi$ and $F=dA$ are the
Berry gauge potential and the associated field strength, respectively,
and $\psi (k)$ is the 2M(k)-dimensional ground state multiplet. The
lattice analog of Eq. (\ref{FuKane}) is \cite{FH}
\begin{alignat}{1}
D_{\rm L} &\equiv\frac{1}{2\pi i}\left[ \sum_{k \in\partial{\cal
B}^-}A(k) -\sum_{k \in{\cal B}^-}F(k) \right]
\nonumber\\
&=-\sum_{k \in{\cal B}^-}n(k)\,\,\, {\text{\rm mod 2}} \, ,
\label{FuHat}
\end{alignat}
since $\sum_{k \in {\cal B}^-}F(k)= \sum_{k \in\partial{\cal
B}^-}A(k) +2\pi i\sum_{k \in{\cal B}^-}n(k)$. $n(k)$ in this
equation are integers and $n(k) \,{\rm mod \, 2} \in \mathbb{Z}_2$
due to the residual $U(1)$ invariance \cite{FH}.

From Eq. (\ref{FuHat}) we infer that the reason for cancelation
of the contribution made by bulk spectrum states is the compactness
of the lattice gauge theory. The existence of the second Dirac cone
points out the nontrivial value $D_{\rm L}=1 \, {\rm {mod\,2}}$ of
the $\mathbb{Z}_2$ invariant in this topologically protected state.

\begin{figure}
\includegraphics[width=.88\columnwidth]{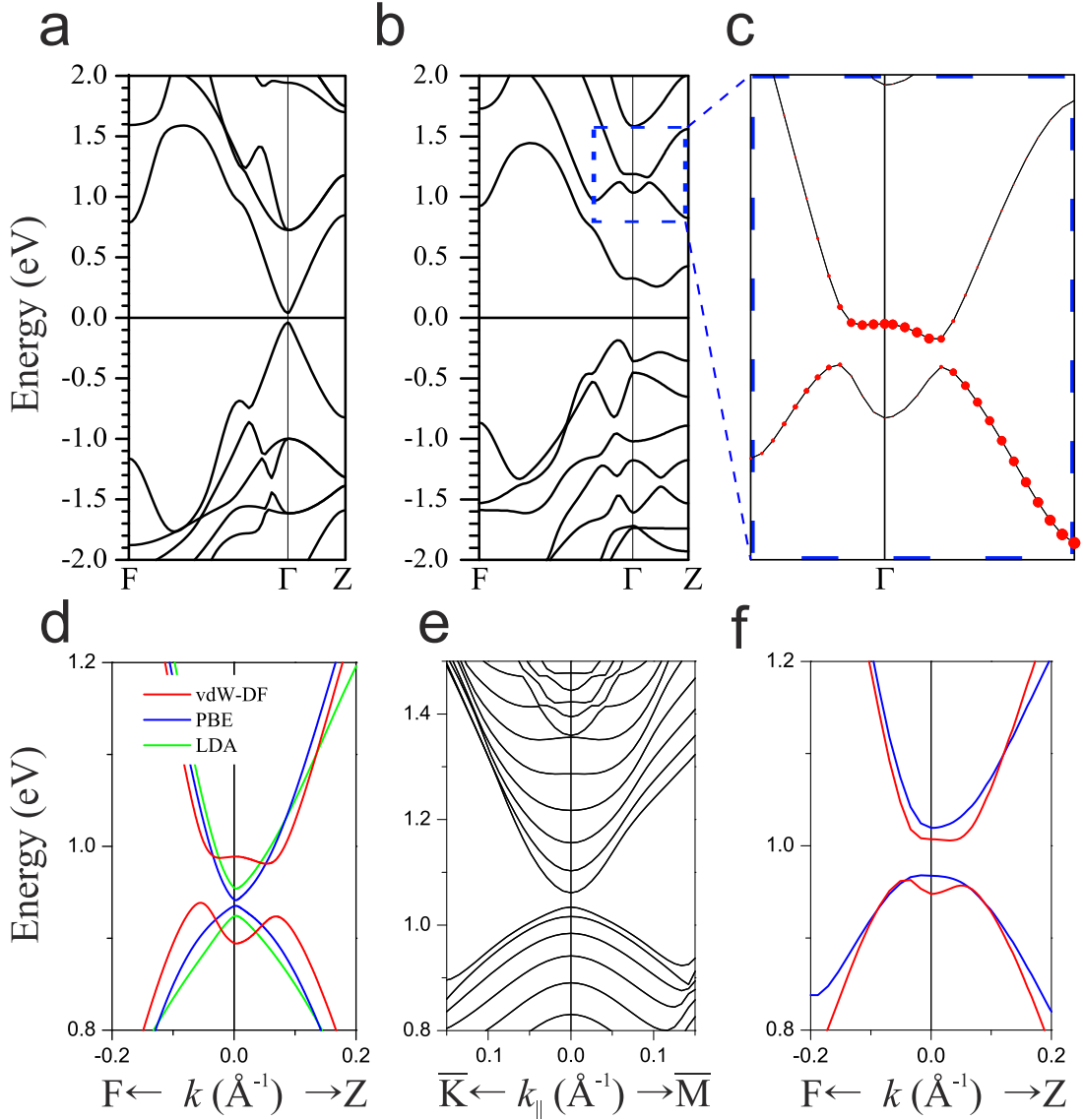}
\caption{\label{fig3}(Color online) Bulk band structure of
Bi$_2$Te$_2$Se {calculated with VASP} (only ${\rm F}-\Gamma$ and
$\Gamma-{\rm Z}$ directions are shown):  (a) without and (b) with
SOI included; (c) magnified view of blue dashed frame marked in
panel (b) with the weight of Se states marked in the second and
third conduction bands near the $\Gamma$ point. {Bulk spectrum
of Bi$_2$Te$_3$ in the vicinity of the local gap within vdW-DF, PBE, and LDA approaches (d). 9 QL-slab electronic
structure of Bi$_2$Te$_3$ within PBE (e). Bulk spectrum of
Bi$_2$Te$_3$ as calculated by FLEUR code with different experimental
parameters: blue line, Ref.~\cite{Wyckoff}; red line,
Ref.~\cite{Nakajima} (f)}.}
\end{figure}

Let us consider now the bulk conduction band of Bi$_2$Te$_2$Se in
the energy range of interest. Without spin-orbit coupling
the second and third conduction bands are degenerate along
the $\Gamma$Z direction (Fig.~\ref{fig3}(a)). Spin-orbit
interaction lifts this degeneracy and opens a gap between these
bands (Fig.~\ref{fig3}(b)). Thus, the gap supporting the new ``X''
shaped surface state in the conduction band originates from SOI\@.
In contrast to the principal $\Gamma$-gap edges, which are composed of
inverted Bi and Te states, both the upper and lower bands of the
conduction band $\Gamma$-gap are mostly composed of Bi states,
however the Se states contribute to these bands as well.  As one
can see in Fig.~\ref{fig3}(c) in the vicinity of $\Gamma$ there are two
parabolic-like bands, gapped at points, where they are inverted due to
the central Se atom contribution.  This inversion of the local
$\Gamma$-gap edges in the conduction band along with the presence at the
same energy of the gaps at F and L TRIMs of the bulk Brillouin zone
(which are projected onto the $\overline{\mathrm M}$ point of the 2D
Brillouin zone) is responsible for the emergence of the unoccupied
Dirac-like surface state at Bi$_2$Te$_2$Se(111).

As one can see in Figs.~\ref{fig2}(e) and (f), the FLEUR
results show the topological conduction-band surface state arising
in Bi$_2$Se$_3$ and Bi$_2$Te$_3$ that is fully consistent with the
outcome of the experiment. The VASP vdW-DF results are almost the
same: they show $10-20$ meV smaller unoccupied gap and slightly
different position of the surface state band crossing within this
gap. The arising of the unoccupied cone in Bi$_2$Se$_3$ and
Bi$_2$Te$_3$, like in Bi$_2$Te$_2$Se, results from inversion of the
QL central atom states of the SOI-induced $\Gamma$ local gap. The
LDA and PBE calculations performed without taking into account the
vdW forces for Bi$_2$Te$_2$Se and Bi$_2$Se$_3$ give the inverted gap
too, while in the case of Bi$_2$Te$_3$  the SOI-induced gap is not
inverted (Fig.~\ref{fig3}(d) and Ref.~\cite{suppvasp}). The dissimilar character of the gap
in Bi$_2$Te$_3$ in PBE and LDA results in absence of the unoccupied cone in
the surface band structure (Fig.~\ref{fig3}(e)). As mentioned above
the different calculations were accompanied by optimization of the
crystal structure. Thus, vdW-DF optimization for all systems under
consideration changes $a$ and $c$ lattice parameter in the range of
$0.5-1$ \% and $0.1-2$ \% as compared to experimental values,
respectively. At the same time both the full structural optimization and the relaxation of atomic positions do not change the
fractional atomic coordinates more than $10^{-3}$. To determine which factor (XC approximation or structural
parameters) is more important for the change of the gap character we
performed calculations of the bulk electronic structure of
Bi$_2$Te$_3$ with two slightly different experimental parameters
(Refs.~\cite{Wyckoff} and \cite{Nakajima}) with the same
exchange-correlation approximation. As one can see in
Fig.~\ref{fig3}(f) this small variation of the crystal structure
changes the character of the gap. The latter means that the
emergence of the unoccupied Dirac state in Bi$_2$Te$_3$ can be
sensitive to the sample preparation. As shown in
Ref.~\cite{Goltsman} even a small deviation from the stoichiometric
composition in Bi$_2$Te$_3$ results in the change of the lattice parameters.
The data measured in the present experiment demonstrate relatively weak intensity in the cone region (Fig.~\ref{fig2}(c)) which may be related to delicate stability of the unoccupied TSS in Bi$_2$Te$_3$ sample used. On the other
hand the highest intensity observed for Bi$_2$Se$_3$ (Fig. 2(b)) reflects the more
localized character of the TSS which is a consequence of the wider gap in
Bi$_2$Se$_3$ as compared to other materials.

\begin{figure}
\includegraphics[width=.75\columnwidth]{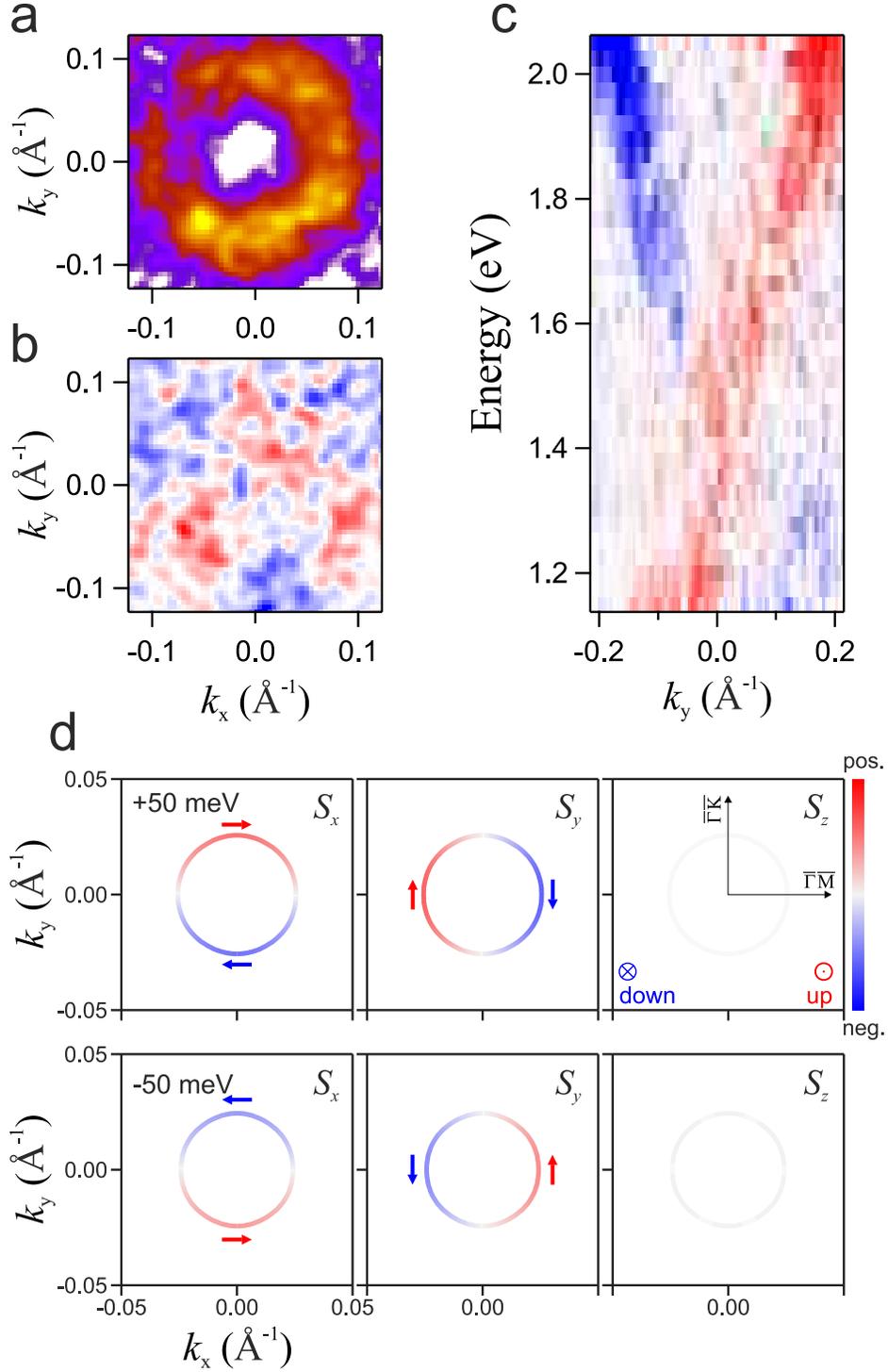}

\caption{\label{fig4}(Color online) MDCs of the sum (a) and
difference (b) of the intensities for left and right polarized light
for an energy of 0.30~eV above the Dirac point for Bi$_2$Se$_3$. (c)
Circular dichroism of unoccupied Dirac cone along the
$\overline\Gamma\overline{\mathrm K}$ direction. {(d) Calculated
spin structure of the unoccupied cone as represented by spin
projections $S_x$, $S_y$, and $S_z$ at an energy of $\pm 50$ meV
with respect to the crossing point.} }
\end{figure}

Additional evidence of the topological character of the unoccupied
surface state emerges from its spin structure which can be accessed
experimentally by use of CD\@.  Therefore MDCs were recorded using
left- and right-handed circularly polarized light.
In Fig.~\ref{fig4}(a) and (b) the sum $I_{\sigma+}$+$I_{\sigma-}$ and the
difference $I_{\sigma+}-I_{\sigma-}$ of the intensities are
shown for Bi$_2$Se$_3$ at an energy 0.3~eV above the Dirac point.
These data were obtained for a different crystal than those in Fig.~\ref{fig2}(b)
which results in a slightly different energy for the Dirac point. 
The normalized asymmetry covers a range of $\pm 12\%$ somewhat smaller
than that observed by ARPES for the occupied Dirac cone where values of
30\% for Bi$_2$Se$_3$ \cite{Park2012} and  up to 40\% for Bi$_2$Te$_3$
\cite{Jung2011,2011arXiv1108.1053S} have been reported. The lower asymmetry in our
experiment might be attributed to the lower angular and energy
resolution than in the ARPES experiments leading to an increased
unpolarized background in the narrow band gap. The two transitions
in 2PPE might also have opposite contributions to the CD pattern
reducing the observed asymmetry.
The MDCs show the expected ring and the CD pattern shows the threefold symmetry
of the substrate as observed further away from the Dirac point for the occupied TSS
\cite{Wang2011,Jung2011,Mirhosseini2012}
Figure~\ref{fig4}(c) demonstrates the development of the CD
pattern over a wide energy range along the
$\overline\Gamma\overline{\mathrm K}$
direction. The dichroism is antisymmetric indicating
opposite spin orientation for the two branches.

Figure~\ref{fig4}(d) presents the
calculated orientation of the electron spin within the surface
state.  The calculated spin-resolved constant energy contours {(CECs)}
below and above the crossing point of the unoccupied cone state show {an ideal circular shape of the CECs in agreement with present experiment and} in-plane spin polarization (out-of-plane component $S_z$ is
negligibly small) with positive (clockwise) spin helicity in the upper
part and negative helicity in the lower part of the cone, i.~e.\ the
same helicity as in the Dirac cone in the principal energy gap.

In summary, we presented a study of the unoccupied part of the band
structure of bismuth chalcogenides.  Empty Dirac cones on
Bi$_2$Se$_3$, Bi$_2$Te$_2$Se, and Bi$_2$Te$_3$ are identified in
both DFT calculations and 2PPE experiments by their linear
dispersion and helical spin structure. Their existence depends on a
symmetry-inverted band gap. {In the case of Bi$_2$Te$_3$ this
inversion is sensitive to small variations of the structural
parameters that can cause dependence of the empty Dirac cone in this
system on sample growth conditions.} The origin of the reduction of
the bulk non-TRIM states contribution has been found using the
$\mathbb{Z}_2$ invariant. The
observed unoccupied TSS opens a new route to measurements on
excitations in and between topological states. For TI-semiconductor
junctions the unoccupied TSS might pave a way to inject and control
spin-polarized currents in the semiconductor conduction band.  The
image-potential state seen at an energy of 4.5~eV in
Fig.~\ref{fig1}(b) corresponds to the situation discussed for
magnetic monopoles \cite{Qi27022009}.  Time-resolved photoemission
experiments on image-potential states and TSSs are in progress.

\begin{acknowledgments}
We thank Philip Hofmann for supplying additional topological
insulator samples showing similar results. SVE thanks Dirk Lamoen
and I.~A. Nechaev for fruitful discussions.
\end{acknowledgments}


\begin{thebibliography}{45}%
\makeatletter
\providecommand \@ifxundefined [1]{%
 \@ifx{#1\undefined}
}%
\providecommand \@ifnum [1]{%
 \ifnum #1\expandafter \@firstoftwo
 \else \expandafter \@secondoftwo
 \fi
}%
\providecommand \@ifx [1]{%
 \ifx #1\expandafter \@firstoftwo
 \else \expandafter \@secondoftwo
 \fi
}%
\providecommand \natexlab [1]{#1}%
\providecommand \enquote  [1]{``#1''}%
\providecommand \bibnamefont  [1]{#1}%
\providecommand \bibfnamefont [1]{#1}%
\providecommand \citenamefont [1]{#1}%
\providecommand \href@noop [0]{\@secondoftwo}%
\providecommand \href [0]{\begingroup \@sanitize@url \@href}%
\providecommand \@href[1]{\@@startlink{#1}\@@href}%
\providecommand \@@href[1]{\endgroup#1\@@endlink}%
\providecommand \@sanitize@url [0]{\catcode `\\12\catcode `\$12\catcode
  `\&12\catcode `\#12\catcode `\^12\catcode `\_12\catcode `\%12\relax}%
\providecommand \@@startlink[1]{}%
\providecommand \@@endlink[0]{}%
\providecommand \url  [0]{\begingroup\@sanitize@url \@url }%
\providecommand \@url [1]{\endgroup\@href {#1}{\urlprefix }}%
\providecommand \urlprefix  [0]{URL }%
\providecommand \Eprint [0]{\href }%
\providecommand \doibase [0]{http://dx.doi.org/}%
\providecommand \selectlanguage [0]{\@gobble}%
\providecommand \bibinfo  [0]{\@secondoftwo}%
\providecommand \bibfield  [0]{\@secondoftwo}%
\providecommand \translation [1]{[#1]}%
\providecommand \BibitemOpen [0]{}%
\providecommand \bibitemStop [0]{}%
\providecommand \bibitemNoStop [0]{.\EOS\space}%
\providecommand \EOS [0]{\spacefactor3000\relax}%
\providecommand \BibitemShut  [1]{\csname bibitem#1\endcsname}%
\let\auto@bib@innerbib\@empty
\bibitem [{\citenamefont {Fu}\ \emph {et~al.}(2007)\citenamefont {Fu},
  \citenamefont {Kane},\ and\ \citenamefont {Mele}}]{FKM_2007}%
  \BibitemOpen
  \bibfield  {author} {\bibinfo {author} {\bibfnamefont {L.}~\bibnamefont
  {Fu}}, \bibinfo {author} {\bibfnamefont {C.~L.}\ \bibnamefont {Kane}}, \ and\
  \bibinfo {author} {\bibfnamefont {E.~J.}\ \bibnamefont {Mele}},\ }\href
  {\doibase 10.1103/PhysRevLett.98.106803} {\bibfield  {journal} {\bibinfo
  {journal} {Phys. Rev. Lett.}\ }\textbf {\bibinfo {volume} {98}},\ \bibinfo
  {pages} {106803} (\bibinfo {year} {2007})}\BibitemShut {NoStop}%
\bibitem [{\citenamefont {Qi}\ \emph {et~al.}(2008)\citenamefont {Qi},
  \citenamefont {Hughes},\ and\ \citenamefont {Zhang}}]{Qi_2008}%
  \BibitemOpen
  \bibfield  {author} {\bibinfo {author} {\bibfnamefont {X.-L.}\ \bibnamefont
  {Qi}}, \bibinfo {author} {\bibfnamefont {T.~L.}\ \bibnamefont {Hughes}}, \
  and\ \bibinfo {author} {\bibfnamefont {S.-C.}\ \bibnamefont {Zhang}},\ }\href
  {\doibase 10.1103/PhysRevB.78.195424} {\bibfield  {journal} {\bibinfo
  {journal} {Phys. Rev. B}\ }\textbf {\bibinfo {volume} {78}},\ \bibinfo
  {pages} {195424} (\bibinfo {year} {2008})}\BibitemShut {NoStop}%
\bibitem [{\citenamefont {Zhang}\ \emph {et~al.}(2009)\citenamefont {Zhang},
  \citenamefont {Liu}, \citenamefont {Qi}, \citenamefont {Dai}, \citenamefont
  {Fang},\ and\ \citenamefont {Zhang}}]{Zhang_NatPhys2009}%
  \BibitemOpen
  \bibfield  {author} {\bibinfo {author} {\bibfnamefont {H.}~\bibnamefont
  {Zhang}}, \bibinfo {author} {\bibfnamefont {C.-X.}\ \bibnamefont {Liu}},
  \bibinfo {author} {\bibfnamefont {X.-L.}\ \bibnamefont {Qi}}, \bibinfo
  {author} {\bibfnamefont {X.}~\bibnamefont {Dai}}, \bibinfo {author}
  {\bibfnamefont {Z.}~\bibnamefont {Fang}}, \ and\ \bibinfo {author}
  {\bibfnamefont {S.-C.}\ \bibnamefont {Zhang}},\ }\href {\doibase
  10.1038/nphys1270} {\bibfield  {journal} {\bibinfo  {journal} {Nature Phys.}\
  }\textbf {\bibinfo {volume} {5}},\ \bibinfo {pages} {438} (\bibinfo {year}
  {2009})}\BibitemShut {NoStop}%
\bibitem [{\citenamefont {Roushan}\ \emph {et~al.}(2009)\citenamefont {Roushan}
  \emph {et~al.}}]{Roushan_Nat2009}%
  \BibitemOpen
  \bibfield  {author} {\bibinfo {author} {\bibfnamefont {P.}~\bibnamefont
  {Roushan}} \emph {et~al.},\ }\href {\doibase 10.1038/nature08308} {\bibfield
  {journal} {\bibinfo  {journal} {Nature}\ }\textbf {\bibinfo {volume} {460}},\
  \bibinfo {pages} {1106} (\bibinfo {year} {2009})}\BibitemShut {NoStop}%
\bibitem [{\citenamefont {Fu}\ and\ \citenamefont {Kane}(2008)}]{FuKane2008}%
  \BibitemOpen
  \bibfield  {author} {\bibinfo {author} {\bibfnamefont {L.}~\bibnamefont
  {Fu}}\ and\ \bibinfo {author} {\bibfnamefont {C.~L.}\ \bibnamefont {Kane}},\
  }\href {\doibase 10.1103/PhysRevLett.100.096407} {\bibfield  {journal}
  {\bibinfo  {journal} {Phys. Rev. Lett.}\ }\textbf {\bibinfo {volume} {100}},\
  \bibinfo {pages} {096407} (\bibinfo {year} {2008})}\BibitemShut {NoStop}%
\bibitem [{\citenamefont {Qi}\ \emph {et~al.}(2009)\citenamefont {Qi},
  \citenamefont {Li}, \citenamefont {Zang},\ and\ \citenamefont
  {Zhang}}]{Qi27022009}%
  \BibitemOpen
  \bibfield  {author} {\bibinfo {author} {\bibfnamefont {X.-L.}\ \bibnamefont
  {Qi}}, \bibinfo {author} {\bibfnamefont {R.}~\bibnamefont {Li}}, \bibinfo
  {author} {\bibfnamefont {J.}~\bibnamefont {Zang}}, \ and\ \bibinfo {author}
  {\bibfnamefont {S.-C.}\ \bibnamefont {Zhang}},\ }\href {\doibase
  10.1126/science.1167747} {\bibfield  {journal} {\bibinfo  {journal}
  {Science}\ }\textbf {\bibinfo {volume} {323}},\ \bibinfo {pages} {1184}
  (\bibinfo {year} {2009})}\BibitemShut {NoStop}%
\bibitem [{\citenamefont {Chen}\ \emph {et~al.}(2009)\citenamefont {Chen} \emph
  {et~al.}}]{Chen10072009}%
  \BibitemOpen
  \bibfield  {author} {\bibinfo {author} {\bibfnamefont {Y.~L.}\ \bibnamefont
  {Chen}} \emph {et~al.},\ }\href {\doibase 10.1126/science.1173034} {\bibfield
   {journal} {\bibinfo  {journal} {Science}\ }\textbf {\bibinfo {volume}
  {325}},\ \bibinfo {pages} {178} (\bibinfo {year} {2009})}\BibitemShut
  {NoStop}%
\bibitem [{\citenamefont {Hsieh}\ \emph {et~al.}(2009)\citenamefont {Hsieh}
  \emph {et~al.}}]{Hsieh09}%
  \BibitemOpen
  \bibfield  {author} {\bibinfo {author} {\bibfnamefont {D.}~\bibnamefont
  {Hsieh}} \emph {et~al.},\ }\href {\doibase 10.1038/nature08234} {\bibfield
  {journal} {\bibinfo  {journal} {Nature}\ }\textbf {\bibinfo {volume} {460}},\
  \bibinfo {pages} {1101} (\bibinfo {year} {2009})}\BibitemShut {NoStop}%
\bibitem [{\citenamefont {Xia}\ \emph {et~al.}(2009)\citenamefont {Xia} \emph
  {et~al.}}]{xia_2009}%
  \BibitemOpen
  \bibfield  {author} {\bibinfo {author} {\bibfnamefont {Y.}~\bibnamefont
  {Xia}} \emph {et~al.},\ }\href {\doibase 10.1038/nphys1274} {\bibfield
  {journal} {\bibinfo  {journal} {Nature Phys.}\ }\textbf {\bibinfo {volume}
  {5}},\ \bibinfo {pages} {398} (\bibinfo {year} {2009})}\BibitemShut {NoStop}%
\bibitem [{\citenamefont {Yazyev}\ \emph {et~al.}(2010)\citenamefont {Yazyev},
  \citenamefont {Moore},\ and\ \citenamefont {Louie}}]{Yazyev2010}%
  \BibitemOpen
  \bibfield  {author} {\bibinfo {author} {\bibfnamefont {O.~V.}\ \bibnamefont
  {Yazyev}}, \bibinfo {author} {\bibfnamefont {J.~E.}\ \bibnamefont {Moore}}, \
  and\ \bibinfo {author} {\bibfnamefont {S.~G.}\ \bibnamefont {Louie}},\ }\href
  {\doibase 10.1103/PhysRevLett.105.266806} {\bibfield  {journal} {\bibinfo
  {journal} {Phys. Rev. Lett.}\ }\textbf {\bibinfo {volume} {105}},\ \bibinfo
  {pages} {266806} (\bibinfo {year} {2010})}\BibitemShut {NoStop}%
\bibitem [{\citenamefont {Eremeev}\ \emph {et~al.}(2010)\citenamefont
  {Eremeev}, \citenamefont {Koroteev},\ and\ \citenamefont
  {Chulkov}}]{EremeevJETPL2010}%
  \BibitemOpen
  \bibfield  {author} {\bibinfo {author} {\bibfnamefont {S.~V.}\ \bibnamefont
  {Eremeev}}, \bibinfo {author} {\bibfnamefont {Y.~M.}\ \bibnamefont
  {Koroteev}}, \ and\ \bibinfo {author} {\bibfnamefont {E.~V.}\ \bibnamefont
  {Chulkov}},\ }\href {\doibase 10.1134/S0021364010080059} {\bibfield
  {journal} {\bibinfo  {journal} {JETP Lett.}\ }\textbf {\bibinfo {volume}
  {91}},\ \bibinfo {pages} {387} (\bibinfo {year} {2010})}\BibitemShut
  {NoStop}%
\bibitem [{\citenamefont {Kuroda}\ \emph {et~al.}(2010)\citenamefont {Kuroda}
  \emph {et~al.}}]{Kuroda2010}%
  \BibitemOpen
  \bibfield  {author} {\bibinfo {author} {\bibfnamefont {K.}~\bibnamefont
  {Kuroda}} \emph {et~al.},\ }\href {\doibase 10.1103/PhysRevLett.105.076802}
  {\bibfield  {journal} {\bibinfo  {journal} {Phys. Rev. Lett.}\ }\textbf
  {\bibinfo {volume} {105}},\ \bibinfo {pages} {076802} (\bibinfo {year}
  {2010})}\BibitemShut {NoStop}%
\bibitem [{\citenamefont {Bianchi}\ \emph {et~al.}(2010)\citenamefont
  {Bianchi}, \citenamefont {Guan}, \citenamefont {Bao}, \citenamefont {Mi},
  \citenamefont {Iversen}, \citenamefont {King},\ and\ \citenamefont
  {Hofmann}}]{bianchi2010}%
  \BibitemOpen
  \bibfield  {author} {\bibinfo {author} {\bibfnamefont {M.}~\bibnamefont
  {Bianchi}}, \bibinfo {author} {\bibfnamefont {D.}~\bibnamefont {Guan}},
  \bibinfo {author} {\bibfnamefont {S.}~\bibnamefont {Bao}}, \bibinfo {author}
  {\bibfnamefont {J.}~\bibnamefont {Mi}}, \bibinfo {author} {\bibfnamefont
  {B.}~\bibnamefont {Iversen}}, \bibinfo {author} {\bibfnamefont
  {P.}~\bibnamefont {King}}, \ and\ \bibinfo {author} {\bibfnamefont
  {P.}~\bibnamefont {Hofmann}},\ }\href {\doibase 10.1038/ncomms1131}
  {\bibfield  {journal} {\bibinfo  {journal} {Nature Commun.}\ }\textbf
  {\bibinfo {volume} {1}},\ \bibinfo {pages} {128} (\bibinfo {year}
  {2010})}\BibitemShut {NoStop}%
\bibitem [{\citenamefont {Wang}\ \emph {et~al.}(2011)\citenamefont {Wang},
  \citenamefont {Hsieh}, \citenamefont {Pilon}, \citenamefont {Fu},
  \citenamefont {Gardner}, \citenamefont {Lee},\ and\ \citenamefont
  {Gedik}}]{Wang2011}%
  \BibitemOpen
  \bibfield  {author} {\bibinfo {author} {\bibfnamefont {Y.~H.}\ \bibnamefont
  {Wang}}, \bibinfo {author} {\bibfnamefont {D.}~\bibnamefont {Hsieh}},
  \bibinfo {author} {\bibfnamefont {D.}~\bibnamefont {Pilon}}, \bibinfo
  {author} {\bibfnamefont {L.}~\bibnamefont {Fu}}, \bibinfo {author}
  {\bibfnamefont {D.~R.}\ \bibnamefont {Gardner}}, \bibinfo {author}
  {\bibfnamefont {Y.~S.}\ \bibnamefont {Lee}}, \ and\ \bibinfo {author}
  {\bibfnamefont {N.}~\bibnamefont {Gedik}},\ }\href {\doibase
  10.1103/PhysRevLett.107.207602} {\bibfield  {journal} {\bibinfo  {journal}
  {Phys. Rev. Lett.}\ }\textbf {\bibinfo {volume} {107}},\ \bibinfo {pages}
  {207602} (\bibinfo {year} {2011})}\BibitemShut {NoStop}%
\bibitem [{\citenamefont {Eremeev}\ \emph {et~al.}(2012)\citenamefont {Eremeev}
  \emph {et~al.}}]{Eremeev_NatComm}%
  \BibitemOpen
  \bibfield  {author} {\bibinfo {author} {\bibfnamefont {S.~V.}\ \bibnamefont
  {Eremeev}} \emph {et~al.},\ }\href {\doibase 10.1038/ncomms1638} {\bibfield
  {journal} {\bibinfo  {journal} {Nature Commun.}\ }\textbf {\bibinfo {volume}
  {3}},\ \bibinfo {pages} {635} (\bibinfo {year} {2012})}\BibitemShut {NoStop}%
\bibitem [{\citenamefont {Ueda}\ \emph {et~al.}(1999)\citenamefont {Ueda},
  \citenamefont {Furuta}, \citenamefont {Okuda}, \citenamefont {Nakatake},
  \citenamefont {Sato}, \citenamefont {Namatame},\ and\ \citenamefont
  {Taniguchi}}]{ueda1999}%
  \BibitemOpen
  \bibfield  {author} {\bibinfo {author} {\bibfnamefont {Y.}~\bibnamefont
  {Ueda}}, \bibinfo {author} {\bibfnamefont {A.}~\bibnamefont {Furuta}},
  \bibinfo {author} {\bibfnamefont {H.}~\bibnamefont {Okuda}}, \bibinfo
  {author} {\bibfnamefont {M.}~\bibnamefont {Nakatake}}, \bibinfo {author}
  {\bibfnamefont {H.}~\bibnamefont {Sato}}, \bibinfo {author} {\bibfnamefont
  {H.}~\bibnamefont {Namatame}}, \ and\ \bibinfo {author} {\bibfnamefont
  {M.}~\bibnamefont {Taniguchi}},\ }\href {\doibase
  10.1016/S0368-2048(98)00335-1} {\bibfield  {journal} {\bibinfo  {journal} {J.
  Electron Spectrosc. Related Phenom.}\ }\textbf {\bibinfo {volume} {101}},\
  \bibinfo {pages} {677} (\bibinfo {year} {1999})}\BibitemShut {NoStop}%
\bibitem [{\citenamefont {Sobota}\ \emph {et~al.}(2012)\citenamefont {Sobota},
  \citenamefont {Yang}, \citenamefont {Analytis}, \citenamefont {Chen},
  \citenamefont {Fisher}, \citenamefont {Kirchmann},\ and\ \citenamefont
  {Shen}}]{PhysRevLett.108.117403}%
  \BibitemOpen
  \bibfield  {author} {\bibinfo {author} {\bibfnamefont {J.~A.}\ \bibnamefont
  {Sobota}}, \bibinfo {author} {\bibfnamefont {S.}~\bibnamefont {Yang}},
  \bibinfo {author} {\bibfnamefont {J.~G.}\ \bibnamefont {Analytis}}, \bibinfo
  {author} {\bibfnamefont {Y.~L.}\ \bibnamefont {Chen}}, \bibinfo {author}
  {\bibfnamefont {I.~R.}\ \bibnamefont {Fisher}}, \bibinfo {author}
  {\bibfnamefont {P.~S.}\ \bibnamefont {Kirchmann}}, \ and\ \bibinfo {author}
  {\bibfnamefont {Z.-X.}\ \bibnamefont {Shen}},\ }\href {\doibase
  10.1103/PhysRevLett.108.117403} {\bibfield  {journal} {\bibinfo  {journal}
  {Phys. Rev. Lett.}\ }\textbf {\bibinfo {volume} {108}},\ \bibinfo {pages}
  {117403} (\bibinfo {year} {2012})}\BibitemShut {NoStop}%
\bibitem [{\citenamefont {Boger}\ \emph {et~al.}(2005)\citenamefont {Boger},
  \citenamefont {{Th.\ Fauster}},\ and\ \citenamefont {Weinelt}}]{Boger05njp}%
  \BibitemOpen
  \bibfield  {author} {\bibinfo {author} {\bibfnamefont {K.}~\bibnamefont
  {Boger}}, \bibinfo {author} {\bibnamefont {{Th.\ Fauster}}}, \ and\ \bibinfo
  {author} {\bibfnamefont {M.}~\bibnamefont {Weinelt}},\ }\href {\doibase
  10.1088/1367-2630/7/1/110} {\bibfield  {journal} {\bibinfo  {journal} {New J.
  Phys.}\ }\textbf {\bibinfo {volume} {7}},\ \bibinfo {pages} {110} (\bibinfo
  {year} {2005})}\BibitemShut {NoStop}%
\bibitem [{sup({\natexlab{a}})}]{suppmat}%
  \BibitemOpen
  \bibinfo {note} {{See Supplemental
  Material at http://link.aps.org/ for details of the photoemission processes
  and the experimental setup.}}\BibitemShut {Stop}%
\bibitem [{\citenamefont {Rieger}\ \emph {et~al.}(1983)\citenamefont {Rieger},
  \citenamefont {Schnell}, \citenamefont {Steinmann},\ and\ \citenamefont
  {Saile}}]{rieg83}%
  \BibitemOpen
  \bibfield  {author} {\bibinfo {author} {\bibfnamefont {D.}~\bibnamefont
  {Rieger}}, \bibinfo {author} {\bibfnamefont {R.~D.}\ \bibnamefont {Schnell}},
  \bibinfo {author} {\bibfnamefont {W.}~\bibnamefont {Steinmann}}, \ and\
  \bibinfo {author} {\bibfnamefont {V.}~\bibnamefont {Saile}},\ }\href
  {\doibase 10.1016/0167-5087(83)91220-6} {\bibfield  {journal} {\bibinfo
  {journal} {Nucl. Instr. Methods}\ }\textbf {\bibinfo {volume} {208}},\
  \bibinfo {pages} {777} (\bibinfo {year} {1983})}\BibitemShut {NoStop}%
\bibitem [{\citenamefont {Kokh}\ \emph {et~al.}(2005)\citenamefont {Kokh},
  \citenamefont {Nenashev}, \citenamefont {Kokh},\ and\ \citenamefont
  {Shvedenkov}}]{Kokh05}%
  \BibitemOpen
  \bibfield  {author} {\bibinfo {author} {\bibfnamefont {K.~A.}\ \bibnamefont
  {Kokh}}, \bibinfo {author} {\bibfnamefont {B.~G.}\ \bibnamefont {Nenashev}},
  \bibinfo {author} {\bibfnamefont {A.~E.}\ \bibnamefont {Kokh}}, \ and\
  \bibinfo {author} {\bibfnamefont {G.~Y.}\ \bibnamefont {Shvedenkov}},\ }\href
  {\doibase 10.1016/j.jcrysgro.2004.11.299} {\bibfield  {journal} {\bibinfo
  {journal} {J. Cryst. Growth}\ }\textbf {\bibinfo {volume} {275}},\ \bibinfo
  {pages} {e2129 } (\bibinfo {year} {2005})}\BibitemShut {NoStop}%
\bibitem [{\citenamefont {Kresse}\ and\ \citenamefont {Hafner}(1993)}]{VASP1}%
  \BibitemOpen
  \bibfield  {author} {\bibinfo {author} {\bibfnamefont {G.}~\bibnamefont
  {Kresse}}\ and\ \bibinfo {author} {\bibfnamefont {J.}~\bibnamefont
  {Hafner}},\ }\href {\doibase 10.1103/PhysRevB.48.13115} {\bibfield  {journal}
  {\bibinfo  {journal} {Phys. Rev. B}\ }\textbf {\bibinfo {volume} {48}},\
  \bibinfo {pages} {13115} (\bibinfo {year} {1993})}\BibitemShut {NoStop}%
\bibitem [{\citenamefont {Kresse}\ and\ \citenamefont
  {Furthm\"uller}(1996)}]{VASP2}%
  \BibitemOpen
  \bibfield  {author} {\bibinfo {author} {\bibfnamefont {G.}~\bibnamefont
  {Kresse}}\ and\ \bibinfo {author} {\bibfnamefont {J.}~\bibnamefont
  {Furthm\"uller}},\ }\href {\doibase 10.1016/0927-0256(96)00008-0} {\bibfield
  {journal} {\bibinfo  {journal} {Comput. Mater. Sci.}\ }\textbf {\bibinfo
  {volume} {6}},\ \bibinfo {pages} {15 } (\bibinfo {year} {1996})}\BibitemShut
  {NoStop}%
\bibitem [{\citenamefont {Bl\"ochl}(1994)}]{PAW1}%
  \BibitemOpen
  \bibfield  {author} {\bibinfo {author} {\bibfnamefont {P.~E.}\ \bibnamefont
  {Bl\"ochl}},\ }\href {\doibase 10.1103/PhysRevB.50.17953} {\bibfield
  {journal} {\bibinfo  {journal} {Phys. Rev. B}\ }\textbf {\bibinfo {volume}
  {50}},\ \bibinfo {pages} {17953} (\bibinfo {year} {1994})}\BibitemShut
  {NoStop}%
\bibitem [{\citenamefont {Kresse}\ and\ \citenamefont {Joubert}(1999)}]{PAW2}%
  \BibitemOpen
  \bibfield  {author} {\bibinfo {author} {\bibfnamefont {G.}~\bibnamefont
  {Kresse}}\ and\ \bibinfo {author} {\bibfnamefont {D.}~\bibnamefont
  {Joubert}},\ }\href {\doibase 10.1103/PhysRevB.59.1758} {\bibfield  {journal}
  {\bibinfo  {journal} {Phys. Rev. B}\ }\textbf {\bibinfo {volume} {59}},\
  \bibinfo {pages} {1758} (\bibinfo {year} {1999})}\BibitemShut {NoStop}%
\bibitem [{\citenamefont {Perdew}\ \emph {et~al.}(1996)\citenamefont {Perdew},
  \citenamefont {Burke},\ and\ \citenamefont {Ernzerhof}}]{PBE}%
  \BibitemOpen
  \bibfield  {author} {\bibinfo {author} {\bibfnamefont {J.~P.}\ \bibnamefont
  {Perdew}}, \bibinfo {author} {\bibfnamefont {K.}~\bibnamefont {Burke}}, \
  and\ \bibinfo {author} {\bibfnamefont {M.}~\bibnamefont {Ernzerhof}},\ }\href
  {\doibase 10.1103/PhysRevLett.77.3865} {\bibfield  {journal} {\bibinfo
  {journal} {Phys. Rev. Lett.}\ }\textbf {\bibinfo {volume} {77}},\ \bibinfo
  {pages} {3865} (\bibinfo {year} {1996})}\BibitemShut {NoStop}%
\bibitem [{\citenamefont {Ceperley}\ and\ \citenamefont {Alder}(1980)}]{LDA}%
  \BibitemOpen
  \bibfield  {author} {\bibinfo {author} {\bibfnamefont {D.~M.}\ \bibnamefont
  {Ceperley}}\ and\ \bibinfo {author} {\bibfnamefont {B.~J.}\ \bibnamefont
  {Alder}},\ }\href {\doibase 10.1103/PhysRevLett.45.566} {\bibfield  {journal}
  {\bibinfo  {journal} {Phys. Rev. Lett.}\ }\textbf {\bibinfo {volume} {45}},\
  \bibinfo {pages} {566} (\bibinfo {year} {1980})}\BibitemShut {NoStop}%
\bibitem [{\citenamefont {Koelling}\ and\ \citenamefont {Harmon}(1977)}]{KH}%
  \BibitemOpen
  \bibfield  {author} {\bibinfo {author} {\bibfnamefont {D.~D.}\ \bibnamefont
  {Koelling}}\ and\ \bibinfo {author} {\bibfnamefont {B.~N.}\ \bibnamefont
  {Harmon}},\ }\href {\doibase 10.1088/0022-3719/10/16/019} {\bibfield
  {journal} {\bibinfo  {journal} {J. Phys. C}\ }\textbf {\bibinfo {volume}
  {10}},\ \bibinfo {pages} {3107} (\bibinfo {year} {1977})}\BibitemShut
  {NoStop}%
\bibitem [{\citenamefont {http://www.flapw.de}()}]{FLAPW}%
  \BibitemOpen
  \bibfield  {author} {\bibinfo {author} {\bibnamefont {http://www.flapw.de}}\
  }\href {http://www.flapw.de} {}
  \BibitemShut
  {Stop}%
\bibitem [{\citenamefont {Li}\ \emph {et~al.}(1990)\citenamefont {Li},
  \citenamefont {Freeman}, \citenamefont {Jansen},\ and\ \citenamefont
  {Fu}}]{Li}%
  \BibitemOpen
  \bibfield  {author} {\bibinfo {author} {\bibfnamefont {C.}~\bibnamefont
  {Li}}, \bibinfo {author} {\bibfnamefont {A.~J.}\ \bibnamefont {Freeman}},
  \bibinfo {author} {\bibfnamefont {H.~J.~F.}\ \bibnamefont {Jansen}}, \ and\
  \bibinfo {author} {\bibfnamefont {C.~L.}\ \bibnamefont {Fu}},\ }\href
  {\doibase 10.1103/PhysRevB.42.5433} {\bibfield  {journal} {\bibinfo
  {journal} {Phys. Rev. B}\ }\textbf {\bibinfo {volume} {42}},\ \bibinfo
  {pages} {5433} (\bibinfo {year} {1990})}\BibitemShut {NoStop}%
\bibitem [{sup({\natexlab{b}})}]{suppvasp}%
  \BibitemOpen
  \bibinfo {note} {{See Ref.~\cite{suppmat}
  for a detailed comparison between VASP and FLEUR calculations.}}\BibitemShut {Stop}%
\bibitem [{\citenamefont {Klime\v{s}}\ \emph {et~al.}(2011)\citenamefont
  {Klime\v{s}}, \citenamefont {Bowler},\ and\ \citenamefont
  {Michaelides}}]{Klimes}%
  \BibitemOpen
  \bibfield  {author} {\bibinfo {author} {\bibfnamefont {J.}~\bibnamefont
  {Klime\v{s}}}, \bibinfo {author} {\bibfnamefont {D.~R.}\ \bibnamefont
  {Bowler}}, \ and\ \bibinfo {author} {\bibfnamefont {A.}~\bibnamefont
  {Michaelides}},\ }\href {\doibase 10.1103/PhysRevB.83.195131} {\bibfield
  {journal} {\bibinfo  {journal} {Phys. Rev. B}\ }\textbf {\bibinfo {volume}
  {83}},\ \bibinfo {pages} {195131} (\bibinfo {year} {2011})}\BibitemShut
  {NoStop}%
\bibitem [{\citenamefont {Neupane}\ \emph {et~al.}(2012)\citenamefont {Neupane}
  \emph {et~al.}}]{Neupane2012}%
  \BibitemOpen
  \bibfield  {author} {\bibinfo {author} {\bibfnamefont {M.}~\bibnamefont
  {Neupane}} \emph {et~al.},\ }\href {\doibase 10.1103/PhysRevB.85.235406}
  {\bibfield  {journal} {\bibinfo  {journal} {Phys. Rev. B}\ }\textbf {\bibinfo
  {volume} {85}},\ \bibinfo {pages} {235406} (\bibinfo {year}
  {2012})}\BibitemShut {NoStop}%
\bibitem [{\citenamefont {Miyamoto}\ \emph {et~al.}(2012)\citenamefont
  {Miyamoto} \emph {et~al.}}]{Miyamoto2012}%
  \BibitemOpen
  \bibfield  {author} {\bibinfo {author} {\bibfnamefont {K.}~\bibnamefont
  {Miyamoto}} \emph {et~al.},\ }\href@noop {} {\bibfield  {journal} {\bibinfo
  {journal} {Phys. Rev. Lett.}\ }\textbf {\bibinfo {volume} {109}} (\bibinfo
  {year} {2012})}\BibitemShut {NoStop}%
\bibitem [{sup({\natexlab{c}})}]{suppmdcs}%
  \BibitemOpen
  \bibinfo {note} {{See Ref.~\cite{suppmat}
  for momentum distribution curves from Bi$_2$Se$_3$.}}\BibitemShut {Stop}%
\bibitem [{\citenamefont {Wang}\ \emph {et~al.}(2010)\citenamefont {Wang},
  \citenamefont {Qi},\ and\ \citenamefont {Zhang}}]{WQZ}%
  \BibitemOpen
  \bibfield  {author} {\bibinfo {author} {\bibfnamefont {Z.}~\bibnamefont
  {Wang}}, \bibinfo {author} {\bibfnamefont {X.-L.}\ \bibnamefont {Qi}}, \ and\
  \bibinfo {author} {\bibfnamefont {S.-C.}\ \bibnamefont {Zhang}},\ }\href
  {\doibase 10.1088/1367-2630/12/6/065007} {\bibfield  {journal} {\bibinfo
  {journal} {New J. Phys.}\ }\textbf {\bibinfo {volume} {12}},\ \bibinfo
  {pages} {065007} (\bibinfo {year} {2010})}\BibitemShut {NoStop}%
\bibitem [{\citenamefont {Fu}\ and\ \citenamefont {Kane}(2006)}]{FK}%
  \BibitemOpen
  \bibfield  {author} {\bibinfo {author} {\bibfnamefont {L.}~\bibnamefont
  {Fu}}\ and\ \bibinfo {author} {\bibfnamefont {C.~L.}\ \bibnamefont {Kane}},\
  }\href {\doibase 10.1103/PhysRevB.74.195312} {\bibfield  {journal} {\bibinfo
  {journal} {Phys. Rev. B}\ }\textbf {\bibinfo {volume} {74}},\ \bibinfo
  {pages} {195312} (\bibinfo {year} {2006})}\BibitemShut {NoStop}%
\bibitem [{\citenamefont {Fukui}\ and\ \citenamefont {Hatsugai}(2007)}]{FH}%
  \BibitemOpen
  \bibfield  {author} {\bibinfo {author} {\bibfnamefont {T.}~\bibnamefont
  {Fukui}}\ and\ \bibinfo {author} {\bibfnamefont {Y.}~\bibnamefont
  {Hatsugai}},\ }\href {\doibase 10.1143/JPSJ.76.053702} {\bibfield  {journal}
  {\bibinfo  {journal} {J. Phys. Soc. Jpn.}\ }\textbf {\bibinfo {volume}
  {76}},\ \bibinfo {pages} {053702} (\bibinfo {year} {2007})}\BibitemShut
  {NoStop}%
\bibitem [{\citenamefont {Wyckoff}(1964)}]{Wyckoff}%
  \BibitemOpen
  \bibfield  {author} {\bibinfo {author} {\bibfnamefont {R.~W.~G.}\
  \bibnamefont {Wyckoff}},\ }\href@noop {} {\emph {\bibinfo {title} {Crystal
  Structures, Vol. 2.}}}\ (\bibinfo  {publisher} {J. Wiley and Sons, New
  York},\ \bibinfo {year} {1964})\BibitemShut {NoStop}%
\bibitem [{\citenamefont {Nakajima}(1963)}]{Nakajima}%
  \BibitemOpen
  \bibfield  {author} {\bibinfo {author} {\bibfnamefont {S.}~\bibnamefont
  {Nakajima}},\ }\href {\doibase 10.1016/0022-3697(63)90207-5} {\bibfield
  {journal} {\bibinfo  {journal} {J. Phys. Chem. Solids}\ }\textbf {\bibinfo
  {volume} {24}},\ \bibinfo {pages} {479} (\bibinfo {year} {1963})}\BibitemShut
  {NoStop}%
\bibitem [{\citenamefont {Goltsman}\ \emph {et~al.}(1973)\citenamefont
  {Goltsman}, \citenamefont {Kudinov},\ and\ \citenamefont
  {Smirnov}}]{Goltsman}%
  \BibitemOpen
  \bibfield  {author} {\bibinfo {author} {\bibfnamefont {B.~M.}\ \bibnamefont
  {Goltsman}}, \bibinfo {author} {\bibfnamefont {B.~A.}\ \bibnamefont
  {Kudinov}}, \ and\ \bibinfo {author} {\bibfnamefont {I.~A.}\ \bibnamefont
  {Smirnov}},\ }\href {http://www.ntis.gov/search/product.aspx?ABBR=AD783734}
  {\emph {\bibinfo {title} {Thermoelectric Semiconductor Materials Based on
  Bi$_2$Te$_3$,}}}\ (\bibinfo  {publisher} {Army Foreign Science \& Technology
  Center},\ \bibinfo {address} {Charlottesville VA},\ \bibinfo {year} {1973})\
  \bibinfo {note} {{Report FSTC-HT-23-1782-73}}\BibitemShut {NoStop}%
\bibitem [{\citenamefont {Park}\ \emph {et~al.}(2012)\citenamefont {Park} \emph
  {et~al.}}]{Park2012}%
  \BibitemOpen
  \bibfield  {author} {\bibinfo {author} {\bibfnamefont {S.~R.}\ \bibnamefont
  {Park}} \emph {et~al.},\ }\href {\doibase 10.1103/PhysRevLett.108.046805}
  {\bibfield  {journal} {\bibinfo  {journal} {Phys. Rev. Lett.}\ }\textbf
  {\bibinfo {volume} {108}},\ \bibinfo {pages} {046805} (\bibinfo {year}
  {2012})}\BibitemShut {NoStop}%
\bibitem [{\citenamefont {Jung}\ \emph {et~al.}(2011)\citenamefont {Jung} \emph
  {et~al.}}]{Jung2011}%
  \BibitemOpen
  \bibfield  {author} {\bibinfo {author} {\bibfnamefont {W.}~\bibnamefont
  {Jung}} \emph {et~al.},\ }\href {\doibase 10.1103/PhysRevB.84.245435}
  {\bibfield  {journal} {\bibinfo  {journal} {Phys. Rev. B}\ }\textbf {\bibinfo
  {volume} {84}},\ \bibinfo {pages} {245435} (\bibinfo {year}
  {2011})}\BibitemShut {NoStop}%
\bibitem [{\citenamefont {{Scholz}}\ \emph {et~al.}()\citenamefont {{Scholz}},
  \citenamefont {{S\'anchez-Barriga}}, \citenamefont {{Marchenko}},
  \citenamefont {{Varykhalov}}, \citenamefont {{Volykhov}}, \citenamefont
  {{Yashina}},\ and\ \citenamefont {{Rader}}}]{2011arXiv1108.1053S}%
  \BibitemOpen
  \bibfield  {author} {\bibinfo {author} {\bibfnamefont {M.~R.}\ \bibnamefont
  {{Scholz}}}, \bibinfo {author} {\bibfnamefont {J.}~\bibnamefont
  {{S\'anchez-Barriga}}}, \bibinfo {author} {\bibfnamefont {D.}~\bibnamefont
  {{Marchenko}}}, \bibinfo {author} {\bibfnamefont {A.}~\bibnamefont
  {{Varykhalov}}}, \bibinfo {author} {\bibfnamefont {A.}~\bibnamefont
  {{Volykhov}}}, \bibinfo {author} {\bibfnamefont {L.~V.}\ \bibnamefont
  {{Yashina}}}, \ and\ \bibinfo {author} {\bibfnamefont {O.}~\bibnamefont
  {{Rader}}},\ }\href {http://arxiv.org/abs/1108.1053} {\bibinfo  {journal}
  {arXiv}\ \bibinfo {pages} {1108.1053}}\BibitemShut {NoStop}%
\bibitem [{\citenamefont {Mirhosseini}\ and\ \citenamefont
  {Henk}(2012)}]{Mirhosseini2012}%
  \BibitemOpen
\bibfield  {journal} {  }\bibfield  {author} {\bibinfo {author} {\bibfnamefont
  {H.}~\bibnamefont {Mirhosseini}}\ and\ \bibinfo {author} {\bibfnamefont
  {J.}~\bibnamefont {Henk}},\ }\href {\doibase 10.1103/PhysRevLett.109.036803}
  {\bibfield  {journal} {\bibinfo  {journal} {Phys. Rev. Lett.}\ }\textbf
  {\bibinfo {volume} {109}},\ \bibinfo {pages} {036803} (\bibinfo {year}
  {2012})}\BibitemShut {NoStop}%
\end{thebibliography}

%

\end{document}